\documentclass[12pt]{article}
\usepackage[english]{babel}
\usepackage{graphicx,graphics}
\newcommand {\vT} {v _ {\scriptscriptstyle T}}
\newcommand {\di} {\displaystyle}
\newcommand {\p} {\partial}

\begin{document}
\title{ Fluid dynamics at arbitrary Knudsen on a base of Alexeev-Boltzmann
  equation: sound in a rarefied gas}
\author{ Leble S. B., Solovchuk M. A., \\
\small   Theoretical Physics and Mathematical Methods Department,\\
\small  Technical University of Gdansk, ul, Narutowicza 11/12,
Gdansk,  Poland,\\
\small  leble@mifgate.pg.gda.pl \\
\small Theoretical Physics Department, \small Immanuel Kant State
University of Russia, Russia, \\
\small 236041, Kaliningrad, Al. Nevsky str. 14. \\
\small  solovchuk@yandex.ru \\  \\[2ex] }
\maketitle

\begin{abstract}
The system of hydrodynamic-type equations is derived from
Alexeev's generalized Boltzmann kinetic equation by two-side
distribution function for a stratified gas in gravity field. It is
applied to a problem of ultrasound propagation and attenuation.
The linearized version of the obtained system is studied and
compared with the Navier-Stokes one at arbitrary Knudsen numbers.
The problem of a generation by a moving plane in a rarefied gas is
explored and used as a test while compared with experiment. It is
good agreement between predicted propagation speed, attenuation
factor and experimental results for a wide range of Knudsen
numbers
\end{abstract}

\section*{Introduction}

Fluid mechanics equations in its' most popular form (Euler,
Navier-Stokes, Fourier-Kirchhoff,Burnett, etc) appear in methods
based on Boltzmann kinetic equation by means of expansion in
Knudsen number(Kn). The first important version of such theory was
made by Hilbert. It implies analiticity in Kn of both distribution
function as well as momenta functions. Further development of the
theory by Chapman-Enskog and Grad \cite{grad1949} weaken
analiticity condition of the momenta on Kn. It allowed to deride
NS, Burnett equations for the Chapman-Enskog method and 13 momenta
Grad equations widely used in fluid dynamics description. Failures
in deep Knudsen regime penetration recovered by direct attempts
with many-moment theories lead to more deep understanding of the
problem \cite{Chapmann1970,Bird,Bobylev}. The Knudsen independent
expansion of the basic (Boltzmann) equation was used, namely one
of Gross-Jackson, starting from the celebrating BGK model. The
unification of Chapman-Enskog and Gross-Jackson
approaches\cite{leble1993} exploits an idea of nonsingular
perturbation method in its Frechet expansion form
\cite{Leble1990}.

One of important verification of fluid dynamics system relates to
the problem of sound propagation. Its simplest version considers
the plane harmonic wave with the correspondent dispersion
relation. Such case obtained by linearization of the basic system
reproduces the known experiments of  \cite{Meyer} rather well. It
incorporated in a direct scheme of kinetic approach
\cite{loyalka}. The fluid mechanics systems, based on BGK
\cite{Gross1959} model of collision integral, obtained recently in
\cite{leble1993} and \cite{Spiegel}, give good results for
velocity of sound in Kn $0.1 \div 10$ but fail in attenuation
description \cite{VLS}.

Developing the method based on Gross-Jackson collision integral
for a non-isotropic fluid for a problem, which specifies
\cite{Lees} a direction in it we use an idea of von Karman to
divide the phase speed with respect of particle velocity direction
along/against the direction axis \cite{VL96}. Such situation takes
place if a gas is stratified in gravity field, that yields
appearance of interne gravity waves branch with the obvious
necessity to account wide range of Kn \cite{VerLeb2005}.

Struchtrup \cite{Struchtrup} regularizes 13-moment Grad equations
doing the same thing as a test. His linearization results in a
dispersion relation, which acoustic branch gives an attenuation
coefficient that also does not fit experiments.

Our article is devoted to this problem; we tried to improve the
results on a way of next Gross-Jackson model \cite{arxiv_LS2006},
the tendency was good but the changes were not enough. Considering
an alternative possibility to compensate the discrepancy in
relaxation timeestimation, we adress to the Alexeev generalization
of Boltzmann equation \cite{Alexeev2004}.

Alexeev-Boltzmann equation looks like:
\begin{equation}
\label{q1} \frac{D f}{D t}-\frac{D }{D t} \left(\tau\frac{D f}{D
t}\right)=J^B \ ,
\end{equation}
where $\frac{D }{D t}= \frac{\p }{\p t}+\vec V\frac{\p }{\p \vec
r}+\frac{\vec F}{m} \frac{\p }{\p \vec V}$ is the
substantional(particle) derivative, $\vec V$  and $\vec r$   are
the velocity and radius vector of the particle, respectively,
$\tau$ is the mean time between collisions, $J^B$ is the collision
Boltzmann integral.

We apply our method for the generalized Boltzmann equation of
Alexeev  and such "joint" theory gives a better agreement with the
experimental data \cite{Meyer} for attenuation at arbitrary
Knudsen number.

\section*{Generalized fluid dynamics equations}

Consider the kinetic equation with the model integral of
collisions in BGK form \cite{Gross1959}:
\begin{equation}
\label{q2}J^B= \nu\left(f_{\it l}-f\right),\
\end{equation}
here $$ f_{\it l}=\di \frac{n}{\pi^{3/2}\vT^3}
\exp\left(-\frac{(\vec V-\vec U)^2}{\vT^2}\right)
$$
- local- equilibrium distribution function. $\vT=\sqrt{2kT/m}$
denotes the average thermal velocity of particles of gas,
$\nu=\nu_0\exp(-z/H)$ -- is the effective frequency of collisions
between particles of the gas at height $z$, $H=kT/mg$ -- is a
parameter of the gas stratification. It is supposed, that density
of the gas n, its average speed $\vec U =(U_x,U_y,U_z)$ and
temperature $T$ are functions of time and coordinates.

Following the idea of the method of piecewise continuous
distribution functions let's search for the solution  $f$  of the
equations(\ref {q1}) as a combination of two locally equilibrium
distribution functions, each of which gives the contribution in
its own area of velocities space:
\begin{equation}
\label{q3} f(t, \vec r, \vec V)=\left\{
\begin{array}{rcl}
 f^+&=&\di n^+ \left(\frac{m}{2\pi k T^+}\right)^{3/2}
\exp\left(-\frac{m(\vec V-\vec U^+)^2}{2 k T^+}\right)\ , \qquad
v_z\ge 0 \vspace{1mm}\\
  f^-&=&\di n^- \left(\frac{m}{2\pi k T^-}\right)^{3/2}
\exp\left(-\frac{m(\vec V-\vec U^-)^2}{2 k T^-}\right)\ ,
\qquad v_z< 0\\
\end{array}
\right.
\end{equation}
here $n^{\pm}, U^{\pm}, T^{\pm}$ depending on $t,z$ are functional
parameters.

The double number of parameters of the distribution function
results in its deviations from a local-equilibrium one. In the
range of small Knudsen numbers $Kn<<1$ we should have
$n^+=n^-$,$T^+=T^-$, $U^+=U^-$ and distribution function start
from a local equilibrium and at the small difference between the
functional 'up' and 'down' parameters produces the Navier-Stokes
equations. The theory is also valid at big Kn(free molecular
regime)\cite{SharipovASA}.

We restrict ourselves by the case of one-dimensional disturbances
$\vec U = (0, 0, U)$, using a set of linearly independent momenta
functions:
 \begin{equation}
\label{q4}
\begin{array}{rclrrcl}
 \varphi_1&=&m\ ,
 & \varphi_4&=& m V_z^2\ ,\vspace{2mm} \\
 \varphi_2&=&m V_z\ ,
 & \varphi_5&=&\di  m V_z V^2\ ,\vspace{2mm}\\
 \varphi_3&=&\di \frac{1}{2}m V^2\ ,
 & \varphi_6&=&\di  m V_z^3\ .\\
\end{array}
\end{equation}

Here the first three functions are collisional invariants. Let's
define a scalar product in velocity space:
\begin{equation}
\label{q5}
 <\varphi_n,f> \equiv <\varphi_n>
 \equiv \int d\vec v\: \varphi_n f\ .
\end{equation}
\begin{equation}
\label{q6}
\begin{array}{lll}
 <m> = \rho \ ,
  &  <m V_z>= \rho U \ ,
 & < \frac{1}{2}m \xi^2> =\frac 32 \frac {\rho}{m}k T \ , \\
<m \xi_z^2> =  P_{zz}\ , & <\frac{1}{2} m \xi_z \xi^2>  = q_z, \
 & <\frac{1}{2} m\xi_z^3> = \di  \bar q_z .\\
\end{array}
\end{equation}
where $ \vec \xi =\vec V - \vec U $ is the peculiar velocity. Here
$\rho=nm$ is mass density, $P_{zz}$ is the diagonal component of
the pressure tensor, $q_z$ is a vertical component of a heat flux
vector, $\bar q_z$ is a parameter having dimension of the heat
flux.

If we now multiply the kinetic equation with the model integral of
collisions in BGK form by $\varphi_i$ and integrate over velocity
space, the fluid dynamic equations appear:

\begin{equation}
\label{q7}
\begin{array}{l}
 \di
 \frac{\p}{\p t}\rho + \frac{\p}{\p z}(\rho U) -\tau \frac{\p^2}{\p^2 t}\rho -
 2\tau \frac{\p^2}{\p t\p z}(\rho U)-\tau \frac{\p^2}{\p^2 z}(P_{zz}+\rho U^2)= 0
\ ,\vspace{2mm} \\
 \di \frac{\p}{\p t}\rho U + \frac{\p}{\p z}(P_{zz}+\rho U^2) -\tau \frac{\p^2}{\p^2 t}(\rho U) -
 2\tau \frac{\p^2}{\p t\p z}(P_{zz}+\rho U^2)
  - \tau \frac{\p^2}{\p^2 z}(2\bar q_z +3UP_{zz}+\rho
 U^3)=0
\ ,\vspace{2mm} \\
 \di
 \left(\frac{\p}{\p t}-\tau \frac{\p^2}{\p^2 t}\right)(\frac {\rho U^2}{2} + \frac 32 \frac {\rho}{m}kT) +
\left( \frac{\p}{\p z}-2\tau \frac{\p^2}{\p t\p z}\right)(\frac
{\rho U^3}{2}+ U\frac 32 \frac {\rho}{m} kT+U P_{zz}+q_z) -
\ \vspace{1mm} \\
\di - \tau \frac{\p^2}{\p^2 z}\left(\frac {\rho
U^4}{2}+2U(q_z+\bar q_z)+U^2(\frac 32 \frac {\rho}{m} kT+\frac 52
P_{zz})+<\frac m2 \xi_z^2 \xi^2>\right)= 0
\ ,\vspace{2mm} \\
\di
 \left(\frac{\p}{\p t}-\tau \frac{\p^2}{\p^2 t}\right)(\rho
 U^2+P_{zz})+
 \left( \frac{\p}{\p z}-2\tau \frac{\p^2}{\p t\p z}\right)(\rho
 U^3+3P_{zz}U+2 \bar q_z)-
\ \vspace{1mm} \\
\di - \tau \frac{\p^2}{\p^2 z}\left(\rho U^4+8U\bar
q_z+6P_{zz}U^2+< m \xi_z^4>\right)= \nu (\frac {\rho}{m}kT-P_{zz})
\ ,\vspace{2mm} \\
\di
 \left(\frac{\p}{\p t}-\tau \frac{\p^2}{\p^2 t}\right)(\rho
 U^3+2P_{zz}U+3\frac {\rho}{m}kTU+2 q_z)+\left( \frac{\p}{\p z}-2\tau \frac{\p^2}{\p t\p
z}\right)(\rho U^4)+
 \ \vspace{1mm} \\
\di+ \left( \frac{\p}{\p z}-2\tau \frac{\p^2}{\p t\p
z}\right)(4U(q_z+\bar q_z)+U^2( 3 \frac {\rho}{m} kT+ 5 P_{zz})+<
m \xi_z^2 \xi^2>)-
\ \vspace{1mm} \\
\di - \tau \frac{\p^2}{\p^2 z}\left(U^2(6q_z+14\bar
q_z)+2<m\xi_z^4>U+3<m\xi_z^2 \xi^2>U+< m \xi_z^3 \xi^2>\right)-
\ \vspace{1mm} \\
\di-\tau \frac{\p^2}{\p^2 z}(\rho U^5+U^3( 3 \frac {\rho}{m}kT+ 9
P_{zz})) =-2\nu q_z - 2 \nu U(P_{zz}-\frac {\rho}{m}kT) \
\ ,\vspace{2mm} \\
\di
 \left(\frac{\p}{\p t}-\tau \frac{\p^2}{\p^2 t}\right)(\rho
 U^3+3P_{zz}U+2 \bar q_z)+
 \ \vspace{1mm} \\
\di+ \left( \frac{\p}{\p z}-2\tau \frac{\p^2}{\p t\p
z}\right)(\rho U^4+8U\bar q_z+6 P_{zz}U^2+< m \xi_z^4>) -
\ \vspace{1mm} \\
\di - \tau \frac{\p^2}{\p^2 z}\left( \rho U^5+10U^3
P_{zz}+20U^2\bar q_z+5<m\xi_z^4>U+< m \xi_z^5>\right)=
\ \vspace{1mm} \\
\di =-2\nu \bar q_z - 3 \nu U(P_{zz}-\frac {\rho}{m} kT), \
\end{array}
\end{equation}
 where
\begin{equation}
\label{q8}
\begin{array}{ll}
 \di  \quad  J_1 = \frac m2 < \xi_z^2 \xi^2> \ ,
 \quad &
 J_2 = \frac m2 <\xi_z^4> \ , \\
\di J_3 = \frac m2 <\xi_z^5>,
 \quad &
 J_4 = \frac m2 <\xi_z^3 \xi^2> \ . \\
\end{array}
\end{equation}

The system (\ref {q7}) of the equations according to the
derivation scheme is valid at all Kn. To close the description it
is enough to plug the two-side distribution function into (\ref
{q6}), that yields for $n^{\pm}, U^{\pm}, T^{\pm}$ as function of
$\rho, U, T, P_{zz},q_z,\bar q_z$. We base here on an expansion in
small Mach numbers  $M = max| \di \frac {U}{v_T}| \ $, up to the
first order. The values of integrals (\ref {q8}) as functions of
thermodynamic parameters of the system (\ref {q7}) are:
\begin{equation}
\label{q9}
\begin{array}{l}
\di  \quad  J_1 =-\frac 52 \rho (\frac {kT_0}{m})^2+ \frac {11}{4}
{\frac {
kT_0 P_{zz}}{m}}+ \frac 94 (\frac {k}{m})^2 \rho T_0 T \ ,\\
\di J_2 = - \frac 32 \rho (\frac {kT_0}{m})^2 + \frac 94 {\frac {
kT_0  P_{zz}}{m}}+ \frac 34  (\frac {k}{m})^2 \rho T_0 T\ , \\
\di J_3 = 6 \frac {kT_0 }{m} \bar q_z + \frac {kT_{0}}{m}q_z +
4\rho_0 U \frac {k^2 T_0^2}{m^2} - 4 \rho U \frac {k^2 T_0^2}{m^2}
\ ,\\
\di J_4 = 6 \frac {kT_0 }{m} \bar q_z + 3\frac {kT_{0}}{m}q_z +
6\rho_0 U \frac {k^2 T_0^2}{m^2} - 6 \rho U \frac {k^2 T_0^2}{m^2}
.
 \
\end{array}
\end{equation}

Substitute $J_I$ into (\ref {q7}) gives modification of fluid
dynamics equations at arbitrary Knudsen numbers.

Let us report the results of investigation of quasi-plane waves
parameters as a function of Kn. For this purpose we proceed in a
standart way:linearizing the system (\ref {q7}) that impose the
dispersion relation as a link between frequency and complex wave
number.

In the model of hard spheres in the continual limit $\tau$ can be
connected with the dynamical viscosity $\eta$ \cite{Chapmann1970}
$$ \tau p = 0.786 \eta.$$
$\tau$ and $\nu$ are linked :
$$ \tau  = 0.786 \nu.$$

\begin{figure}
  \includegraphics[ height=9cm]{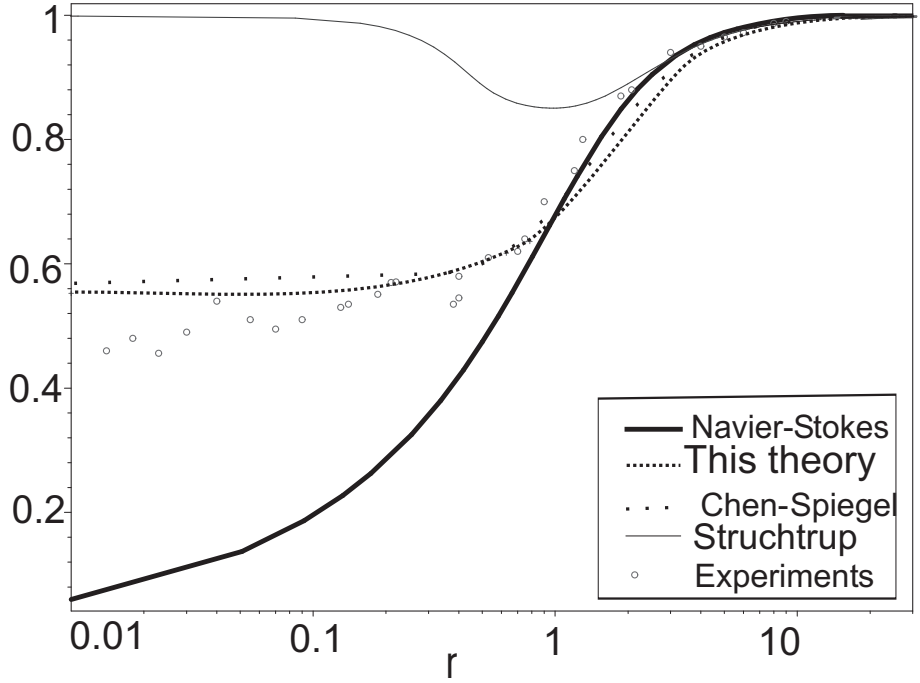}
  \caption{The inverse non-dimensional phase velocity as a
function of the inverse Knudsen number. The results of this paper
are compared to Navier-Stokes, Chen-Spiegel \cite{Spiegel},
regularization of Grad's method \cite{Struchtrup} and the
experimental data \cite{Meyer}}
 \end{figure}
\begin{figure}
  \includegraphics[height=9cm]{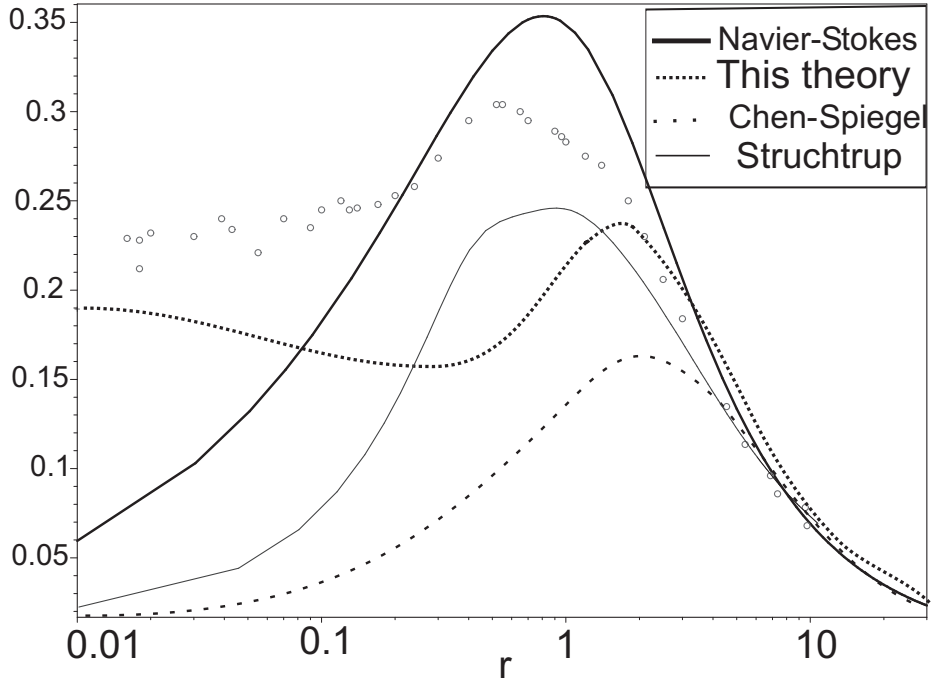}
  \caption{The attenuation factor of the linear disturbance as a
function of the inverse Knudsen number.}
 \end{figure}

In figures 1, 2 a comparison of our results of numerical
calculation of dimensionless sound speed and attenuation factor
depending on $r \sim \frac {1}{Kn}$  is carried out in a parallel
way with the results by other authors. The Navier-Stokes
prediction is qualitatively wrong at big Knudsen number. Our
results for phase speed give the good consistency with the
experiments at all Knudsen numbers. However, our results for the
attenuation of ultrasound are good (as we can see in experiment)
for the number r up to order unity and in the free molecular
regime. Taking into account disadvantages of model integral of
collisions it is planed to consider kinetic equation with full
integral of collisions. It will permit to describe processes in
transition regime.

\bibliographystyle{aipprocl} 

\end{document}